\documentclass[twocolumn,showpacs,
  superscriptaddress,
  notitlepage,showkeys,
  nofootinbib]{aastex631}
\usepackage{amssymb}
\usepackage{amsmath}
\usepackage{graphicx}
\usepackage{dcolumn}
\usepackage{hyperref}
\usepackage{color,units}
\usepackage{aas_macros}
\usepackage{xspace}
\usepackage[normalem]{ulem} 
\usepackage{soul}

\newcommand{\SPA}{School of Physics and Astronomy, Monash University, Vic 3800, Australia}
\newcommand{\OzGravMonash}{OzGrav: The ARC Centre of Excellence for Gravitational Wave Discovery, Clayton VIC 3800, Australia}

\newcommand{\bilby}{\texttt{Bilby}}
\newcommand{\tbilby}{\texttt{tBilby}}

\begin{document}

\title{Transdimensional inference for gravitational-wave astronomy with \bilby}

\author{Hui Tong}
\affiliation{\SPA}
\affiliation{\OzGravMonash}
\correspondingauthor{Hui Tong}
\email{hui.tong@monash.edu}

\author{Nir Guttman}
\affiliation{\SPA}
\affiliation{\OzGravMonash}

\author{Teagan A. Clarke}
\affiliation{\SPA}
\affiliation{\OzGravMonash}

\author{Paul D. Lasky}
\affiliation{\SPA}
\affiliation{\OzGravMonash}

\author{Eric Thrane}
\affiliation{\SPA}
\affiliation{\OzGravMonash}

\author{Ethan Payne}
\affiliation{Department of Physics, California Institute of Technology, Pasadena, California 91125, USA}
\affiliation{LIGO Laboratory, California Institute of Technology, Pasadena, California 91125, USA}

\author{Rowina Nathan}
\affiliation{\SPA}
\affiliation{\OzGravMonash}

\author{Ben Farr}
\affiliation{Institute for Fundamental Science, Department of Physics, University of Oregon, Eugene, OR 97403, USA}

\author{Maya Fishbach}
\affiliation{Canadian Institute for Theoretical Astrophysics, 60 St George St, University of Toronto, Toronto, ON M5S 3H8, Canada}
\affiliation{David A. Dunlap Department of Astronomy and Astrophysics, 50 St George St, University of Toronto, Toronto, ON M5S 3H8, Canada}
\affiliation{Department of Physics, 60 St George St, University of Toronto, Toronto, ON M5S 3H8, Canada}

\author{Gregory Ashton}
\affiliation{Department of Physics, Royal Holloway, University of London, TW20 0EX, UK}

\author{Valentina Di Marco}
\affiliation{\SPA}
\affiliation{\OzGravMonash}

\begin{abstract}
It has become increasingly useful to answer questions in gravitational-wave astronomy using \textit{transdimensional} models where the number of free parameters can be varied depending on the complexity required to fit the data.
Given the growing interest in transdimensional inference, we introduce a new package for the Bayesian inference Library (\bilby) called \tbilby. 
The \tbilby{} package allows users to set up transdimensional inference calculations using the existing \bilby{} architecture with off-the-shelf nested samplers and/or Markov Chain Monte Carlo algorithms.
Transdimensional models are particularly helpful when we seek to test theoretically uncertain predictions described by phenomenological models. 
For example, bursts of gravitational waves can be modelled using a superposition of $N$ wavelets where $N$ is itself a free parameter.
Short pulses are modelled with small values of $N$ whereas longer, more complicated signals are represented with a large number of wavelets stitched together.
Other transdimensional models have found use describing instrumental noise and the population properties of gravitational-wave sources.
We provide a few demonstrations of \tbilby{}, including fitting the gravitational-wave signal GW150914 with a superposition of $N$ sine-Gaussian wavelets.
We outline our plans to further develop the \tbilby{} code suite for a broader range of transdimensional problems.
\end{abstract}


\section{Introduction}
Since the first detection of gravitational waves~\citep{gw150914}, Bayesian inference has been widely used to infer the astrophysical properties of merging binaries~\citep{gw150914_properties}.
Bayesian inference is used to search for physics beyond general relativity~\citep{gw150914_tgr}, to probe nuclear physics at extreme densities~\citep{gw170817_eos}, to measure the expansion of the Universe~\citep{gw170817_cosmo,Hotokezaka}, and to study the formation of merging binaries~\citep{o3a_pop,gwtc-3_pop}.

In many applications, the framework underpinning these inferences is theoretically precise; that is, we have trustworthy, quantitative predictions for the data given the model parameters.
For example, when we infer the masses of merging black holes, we are able to leverage state-of-the-art gravitational waveforms, built from numerical-relativity simulations, to interpret data.
In other cases, however, there is significant theoretical uncertainty and so we rely on phenomenological models.
For example, following the detection of GW150914, the LIGO–Virgo Collaborations used the \texttt{BayesWave} algorithm \citep{Cornish_2015} to perform a study to reconstruct the strain time series in the data with minimal assumptions using a superposition of $N$ sine-Gaussian wavelets \citep{gw150914_bw,cwb}.\footnote{Sine-Gaussian functions are sometimes called Morlet or Gabor wavelets~\citep{Morlet}}
If we treat $N$ as a free parameter, then the total number of model parameters is itself variable.
Such an analysis---where the number of free parameters is itself a free parameter---is said to be \textit{transdimensional}.
The striking agreement between LIGO--Virgo's minimal-assumption reconstruction and the waveform predicted by general relativity helped cement the interpretation of the signal as a binary black hole merger \citep{gw150914_bw}.
It remains a powerful demonstration of the usefulness of transdimensional models.

There are other noteworthy applications of transdimensional inference in gravitational-wave astronomy.
In the audio band where the LIGO--Virgo--KAGRA~\cite[LVK;][]{aLIGO, aVIRGO, kagra_2013} observatories operate, the \texttt{BayesWave} package \citep{Cornish_2015,BayesWave2021} has been used for minimum-assumption model checking and waveform reconstruction~\citep{ Millhouse2018,Pannarale2019,Dalya2021}, improving the statistical significance of short and unmodeled ``bursting'' signals~\citep{Littenberg2016,Shuen2021},  modelling astrophysically uncertain waveforms (e.g., from supernovae and hypermassive neutron stars) \citep{Raza,Miravet-Tenes,Ashton_2022}, modelling deviations from general relativity \citep{Chatziioannou2021,Nathan}, and  subtracting noise artifacts (glitches) \citep{Littenberg2010,Pankow2018,Chatziioannou_cbc_glitch, Davis2022, Hourihane_2022}. 
Meanwhile, the related \texttt{BayesLine} code is frequently used to estimate the noise power spectral density of gravitational-wave measurements \citep{BayesLine, Gupta}.
Transdimensional analyses have also been demonstrated for use in the millihertz band by space-based observatories~\citep{Littenberg2020} and in the nanohertz band by pulsar timing arrays~\citep{Ellis}.
The code package \texttt{Eryn}~\citep{Eryn} was recently introduced as a multi-purpose tool for transdimensional inference with special attention to problems relevant for the LVK and LISA.

In this work, we introduce \tbilby, a package for transdimensional sampling with the Bayesian Inference Library \bilby{} \citep{bilby,bilby_gwtc1}.
\bilby{} is widely used in gravitational-wave astronomy.
It is designed and maintained with four guiding principles: modularity, consistency, generality, and usability.
The mission for \bilby{} is to be intuitive enough to be used by new researchers, while still being applicable to a broad class of problems, and with the ability to easily swap samplers when needed.
Our goal is to leverage these attributes, building on the existing \bilby{} infrastructure, in order to make it easier for \bilby{} users to carry out transdimensional analyses.

We envision the \tbilby{} project as a long-term effort that will be developed gradually.
With this in mind, we start here with a specific class of transdimensional problems: transient waveforms that can be modelled with a superposition of $N$ component functions.
In particular, we demonstrate a minimum-assumption reconstruction of GW150914 using a superposition of $N$ sine-Gaussian functions.
We use this demonstration to explain key concepts in transdimensional inference including the notion of ghost parameters and order statistics of a uniform distribution.\footnote{A dedicated appendix is provided for each of these topics: Appendix \ref{appendix:ghost parameter} covers ghost parameters while Appendix \ref{appendix:order stats} discusses the mathematics of order statistics.}
Readers can reproduce our calculation using the accompanying code.\footnote{The code can be found at the \texttt{git} repository \url{https://github.com/tBilby/tBilby}.}
The remainder of this paper is organized as follows.
In Section~\ref{Sec:stat}, we cover the basic principles of transdimensional inference and describe how they are implemented in \tbilby.
In Section~\ref{sec:demonstration}, we demonstrate the \tbilby{} package with two examples: a toy-model problem consisting of a superposition of Gaussian pulses and a minimum-assumption reconstruction of GW150914.
We provide closing remarks in Section~\ref{Sec:discussion}, briefly demonstrating another transdimensional example fitting LIGO's noise amplitude spectral density with a sum of $N$ power laws and $M$ Lorentzian functions. We also sketch our priorities for future development.

\section{Method}\label{Sec:stat}
One of the goals of Bayesian inference is to determine the posterior distribution for model parameters $\vec\theta$ given a prior $\pi(\vec\theta) $, data $d$, and likelihood ${\cal L}(d|\vec\theta)$. 
In a transdimensional problem, the number of parameters $N$ is itself a parameter:
\begin{equation}
    \vec\theta \equiv \{ \theta_1, \ldots,\theta_N, N \}.
\end{equation}
In some cases, this problem can be solved with brute-force parallelization: one can run multiple inference jobs, each with a different \textit{fixed} number of parameters $N$, and then combine the resulting samples based on the Bayesian evidence for each fixed-$N$ analysis ${\cal Z}_N,$ as well as their prior preference.
This approach works adequately when there is a relatively small range of values for $N$. 
However, it becomes inefficient when time is wasted exploring many values of $N$ disfavoured by the likelihood function. 
The solution is to sample in $N$. \footnote{In the context of Markov chain Monte Carlo samplers, this is essentially the same as the Reversible jump Markov chain Monte Carlo technique \citep{Peter_RJMCMC}.}

The number of parameters $N$ is treated similarly to any other discrete parameter in \bilby.
In our demonstrations below, we take the prior $\pi(N)$ to be uniform on the interval $[N_\text{min}, N_\text{max}]$.
At each step, the sampler draws a value of $N$ along with values for all possible parameters in $\vec\theta$---even parameters $\theta_{k>N}$ that are not used for the $N$-parameter model. 
We refer to the $\theta_{k>N}$ parameters as ``ghost parameters'' since they are not included in the likelihood evaluation, similar to the method in \citet{pseudo_prior, Liu_2023}. 
In App.~\ref{appendix:ghost parameter}, we prove that when we marginalize over the ghost parameters, we obtain the same posterior as one would obtain without ghost parameters using either the brute-force method or the transdimensional sampler.\footnote{It is interesting to note that ghost parameters $\theta_{k>N}$ incur no Occam penalty. Since the ghost parameters do not appear in the likelihood, the flexibility of the model has not changed, so there is no penalty for adding unnecessary complexity.}

From the perspective of the \tbilby{} code, the transdimensional model behaves like a fixed-dimensional model in order to obtain the joint posterior:
\begin{equation}
\begin{aligned}   
p(&\theta_1,\ldots \theta_{N_\text{max}}, N \mid d) \propto \\
& \pi(\theta_1,\ldots \theta_{N_\text{max}}) \, \pi(N) \, \mathcal{L}(d| \theta_1,\ldots\theta_N).
\end{aligned}
\end{equation}
Since the likelihood does not depend on the ghost parameters, the marginal posterior distribution for the $k>N$ ghost parameters is equivalent to the prior for the ghost parameters
\begin{equation}
\begin{aligned}
    p(&\theta_{k} | d, N, k>N, \theta_{j \le N}) = \pi(\theta_{k}|\theta_{j\le N}) .
\end{aligned}
\end{equation}
The $k>N$ ghost-parameter samples are removed in post-processing since they are not actually part of our model.
The ghost-parameter framework is convenient since it allows us to perform transdimensional inference using the off-the-shelf samplers already available in \bilby.\footnote{The additional computational cost incurred by drawing prior samples that we do not use is (for most applications that we foresee) negligible compared the cost of the likelihood evaluation.}

In this paper, we mainly focus on a specific set of transdimensional problems in which the data consists of a time series $d(t)$, and the signal model $s(t)$ is modelled with a sum of $N$ components, each with parameters $\theta_k$:
    \begin{equation}
        s(t | \vec\theta) = \sum_{k=1}^N s_k(t | \theta_k) .
    \end{equation}
We refer to each $s_k(t | \theta_k)$ as a \textit{component function}.

An interesting issue arises fitting signals with a superposition of identical component functions \citep{Buscicchio_2019}. 
Since the component functions are identical, the labels of parameters can be swapped without changing the likelihood, leading to multiple likelihood peaks all describing the same fit.
Given $N$ component functions, the number of these degenerate modes scales like $N!$.
Thus, multimodality quickly becomes problematic if left unchecked.
One must therefore intervene to avoid a multimodal likelihood surface that needlessly complicates our sampling efforts.

We address this issue by using the mathematics of \textit{order statistics} \citep{order_stats}. 
The basic idea of order statistics is to devise a method for ranking what would otherwise be indistinguishable component functions so that each one is uniquely identified.
This eliminates the artificial multimodality arising from label swapping.

Of course, there are different ways in which component functions can be ordered: a set of Morlet wavelets, for example, can be ordered by frequency, time, amplitude, etc.
The choice of a suitable ordering strategy is problem-dependent.
However, for many transdimensional problems, it is useful to order component functions by descending signal-to-noise ratio (SNR).

As the sampler explores the likelihood surface, it is likely to find features in the order of each feature’s contribution to likelihood. 
Therefore, a natural way to order components is by descending SNR so that the first component is the one with the highest SNR. 
This way, the order of components are reasonably well matched with the order that the sampler finds the features, which reduces the chances that the sampler becomes stuck in a local likelihood maximum. 
The first component, with the largest SNR, is likely to be detected by the sampler first, followed by the second component with the  second largest SNR component, and so on.
In the next Section, we demonstrate the principles outlined in this Section with examples.

While we highlight the usefulness of SNR ordering, \tbilby{} users are free to implement whichever ordering strategy they choose.
In Appendix~\ref{alternative}, we discuss alternative ordering schemes, showing that they are mathematically equivalent, but that SNR ordering is more reliable for many problems of interest.

\section{Demonstration}\label{sec:demonstration}
\subsection{A superposition of Gaussian pulses}\label{sec:pulses}
As a warm-up exercise, we consider a simple problem of fitting data with $N$ Gaussian pulses in the presence of Gaussian white noise.\footnote{The calculations in this subsection are performed in the accompanying \texttt{jupyter} notebook, \texttt{pulse.ipynb}, in the git repository linked above.}
Our data $d(t)$ is a time series consisting of signal $s(t)$ and noise $n(t)$:
\begin{equation}
    d(t) = s(t) + n(t) .
\end{equation}
Our signal model is a superposition of component functions given by
\begin{equation}
    s(t | \mu_k, A_k, \sigma_k) = \frac{A_k}{\sigma_k \sqrt{2\pi}} \exp\left(-\frac{(t-\mu_k)^2}{2\sigma_k^2} \right).
\end{equation}
The mean $\mu_k$, amplitude $A_k$ and width $\sigma_k$ are free to vary.
We simulated data with a signal consisting of $N=3$ pulses plus Gaussian white noise $n(t)$ with variance
\begin{equation}
    \sigma_n=\sqrt{\langle n(t) n(t') \rangle} = 0.15 \, \delta_{tt'} ,
\end{equation}
where $\delta_{tt'}$ is a Kronecker delta function.

\begin{figure}
\centering\includegraphics{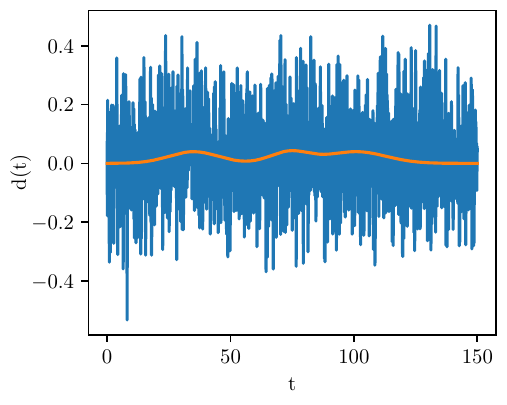}
    \caption{
    Simulated data (blue) for our Gaussian pulse model described in Subsection~\ref{sec:pulses}.
    The signal (orange) consists of three Gaussian pulses.
    }
    \label{fig:gaussian_toy_model}
\end{figure}

The corresponding SNR expression of each Gaussian can be written as
\begin{equation}
    \text{SNR}_k\propto\frac{A_k}{\sqrt{\sigma_k}\sigma_n}.
\end{equation}
The likelihood is
\begin{equation}
    \mathcal{L}(d|s,\sigma_n) = 
    \prod_i 
    \frac{1}{\sqrt{2\pi\sigma_n^2}}\text{exp}\left(\frac{-\Big(d(t_i)-s(t_i)\Big)^2}{2\sigma_n^2}\right),
\end{equation}
where $d$ is the data, $s$ is the signal template, $t_i$ is a discrete time, and $\sigma_n$ is the noise.
We show simulated data in Fig.~\ref{fig:gaussian_toy_model} created with parameters summarized in Table~\ref{tab:toy_model_true}.

\begin{figure}
\centering\includegraphics{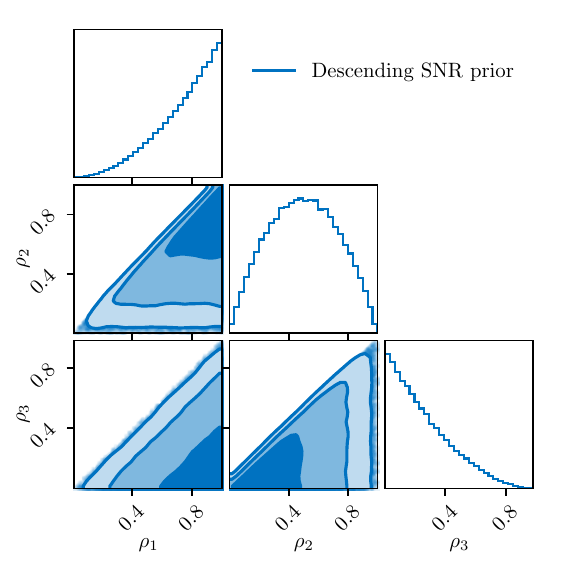}
    \caption{
    An example of descending SNR priors. Note here the variables are relative SNR $\rho_i$ defined in Eq.~\ref{eq:relative_SNR}
    }
    \label{fig:order_stats_prior_exmple}
\end{figure}

Our prior for the number of pulses $N$ is a discrete uniform distribution $\mathcal{U}(0,6)$. 
We order each pulse in descending order according to its SNR.
We assume that the \textit{unordered} SNR values are uniformly distributed on the interval $(0,\text{SNR}_\text{max})$ where $\text{SNR}_\text{max}=10$.
It follows that the prior on the first \textit{ordered} SNR is given by a beta distribution with parameters $\alpha=N$ and $\beta = 1$  \citep{order_stats}:
\begin{align}\label{eq:Beta}
    \pi(\rho_1) \propto \text{Beta}(\rho_1| N, 1) .
\end{align}
Here, $\rho_1$ is the ``relative signal-to-noise ratio'' given by 
\begin{align}\label{eq:relative_SNR}
    \rho_i = \text{SNR}_i / \text{SNR}_\text{max} .
\end{align}
The joint prior for subsequent ordered SNR values when $i<j$ are given by  \citep{order_stats}
\begin{align}\label{eq:joint}
    \pi(\rho_i, \rho_j) = N! 
    \frac{(1-\rho_i^{i-1})}{(i-1)!}
    \frac{(\rho_i - \rho_j)^{j-i-1}}{(j-i-1)!}
    \frac{\rho_j^{N-j}}{(N-j)!}
\end{align}
Using Eqs.~\ref{eq:Beta}-\ref{eq:joint}, one can draw a full set of ordered SNRs.
A corner plot showing an example prior for ordered SNRs is provided in Fig.~\ref{fig:order_stats_prior_exmple}.
Additional details are provided in \ref{appendix:order stats}.

The prior for the central time of the pulse $\mu_{k}$ is uniform over the interval $(0,150)$.
For the widths, we employ a uniform priors $U[5,20]$.

\begin{table}
\centering
 \begin{tabular}{|c| c| c| c|} 
 \hline
 Pulse & Mean & Amplitude & Width\\
 \hline
 1 & 35 & 1.0 & 10 \\
 \hline
 2 & 74 & 0.8 & 8 \\
 \hline
 3 & 101 & 1.2 & 12 \\
 \hline
 \end{tabular}
 \caption{
 Parameter values for the Gaussian pulses shown in Fig.\ref{fig:gaussian_toy_model}.
 }
 \label{tab:toy_model_true}
\end{table}

We run \tbilby{} using two different samplers: the nested sampler \texttt{dynesty} \citep{dynesty} and the parallel-tempered Markov chain Monte Carlo sampler \texttt{ptemcee} \citep{ptemcee}.
For \texttt{ptemcee}, we update the number of pulses $N$ and the parameters for each pulse $\theta_k$ separately every time a new move is proposed in the sampling process.
Since we are using ghost parameters, the sampler behaves as though it is exploring a fixed-dimensional space.
In each iteration, we randomly add a pulse, remove a pulse, or keep the number of pulses fixed with equal probability. 
Since \texttt{dynesty} draws samples from priors, jumps in $N$ occur automatically by virtue of the discrete prior $\pi(N)$.

\begin{figure}
    \centering
    \includegraphics{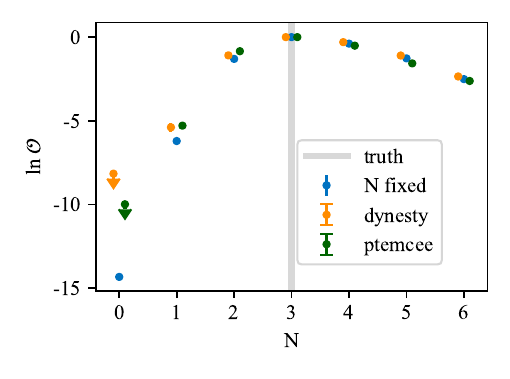}
    \caption{ Natural log posterior odds obtained with different sampling techniques (see Eq. \ref{eq:odds}). The odds are measured relative to the favored $N=3$ model. The uncertainties are one-sigma. The orange and green points are transdimensional sampling results using \texttt{dynesty} and \texttt{ptemcee}. The navy blue points labelled by ``N fixed"  where we calculate the evidence for each value of $N$ with dedicated \texttt{dynesty} runs provide the ground truth.}
    \label{fig:odds}
\end{figure}

In Fig.~\ref{fig:odds}, we plot the posterior odds
\begin{equation}\label{eq:odds}
    {\cal O}(N) = \frac{{\cal L}(d | N)}{{\cal L}(d | N'=3) } ,
\end{equation}
which compares the posterior support for different values of $N$ to the best-fit $N=3$ model.\footnote{Astute readers may notice that the right-hand side of Eq.~\ref{eq:odds} does not include the prior odds. This is because the prior odds in this case are unity.}\footnote{We estimate the uncertainty in our $\ln {\cal O}$ calculations as follows:
\begin{equation}
    \sigma_{\ln \cal O}^2 = \frac{1}{n_N} + \frac{1}{n_{\o}} .
\end{equation}
Here, $n_N$ is the number of posterior samples for the hypothesis that the data are described by $N$ component functions while $n_{\o}$ is the number of posterior samples describing the fiducial model---in this case, $N=3$.}
The results obtained with \texttt{dynesty} are shown in orange while the results obtained with \texttt{ptemcee} are shown in green.
As a sanity check, we also use \texttt{dynesty} to calculate the marginal likelihood for each value of $N$ with separate fixed-$N$, which, combined with our prior of $N$, we use to estimate the ground-truth posterior obtained without transdimensional inference.
All three methods produce a similar distribution.
Some values of $N$ are strongly disfavored, and so the transdimensional sample records no posterior samples for that value of $N$.
In such cases, we set an upper limit on $\ln {\cal O}$.
Both \texttt{dynesty} and \texttt{ptemcee} produce $\ln{\cal O}$ values that are consistent with the fixed-$N$ ground truth. 

We compare the computational cost between the brute-force method of performing many fixed-$N$ runs and using \tbilby. The fixed-$N$ runs for $N\in [0, 6]$ take roughly 5.2 times the sampling time of the tBilby \texttt{dynesty} run with the same sampler settings.\footnote{Note this is not a rigorous apples-to-apples comparison. For example, we do not require the same number of effective samples between the brute-force calculation and \tbilby. 
However, it does provide a rough estimate of the improvement in computational cost for this particular problem.}
As expected, when we run different $N$ models separately, most of the computational time is spent exploring complicated models with large $N$, which may not be the models with the highest Bayes factor. Since transdimensional sampling accounts for the Occam factor during sampling process, it automatically prevents the sampler exploring disfavoured regions of the parameter space.

\begin{figure*}
    \centering
    \includegraphics[width=0.9\textwidth]{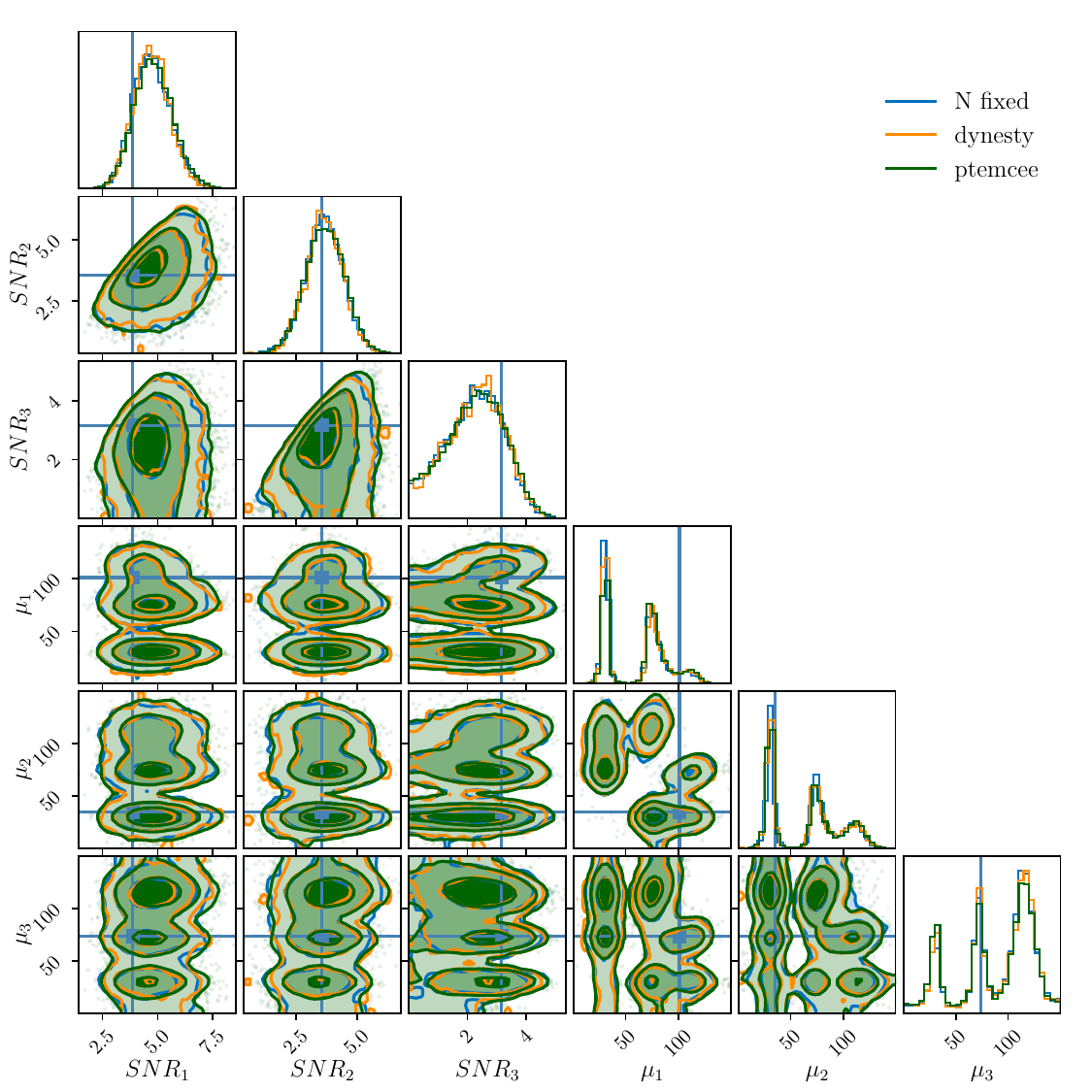}
    \caption{
    Posterior distribution for the SNRs and the means of $N=3$ Gaussian pulses with descending SNR priors.
    The different shades show one, two, and three-sigma credible intervals.
    The ground truth, obtained with a fixed-$N$ \texttt{dynesty} run, is shown in blue.
    In orange we plot the results obtained using a transdimensional implementation of \texttt{dyensty} while green shows a transdimensional implementation of \texttt{ptemcee}.}
    \label{fig:original_posterior_toy}
\end{figure*}

In Fig.~\ref{fig:original_posterior_toy}, we present a corner plot showing the marginal posterior distribution of parameters $\text{SNR}_1,\text{SNR}_2,\text{SNR}_3, \mu_1, \mu_2, \mu_3$ given samples $N=3$.
As above, the fixed-$N$ ground truth is shown in blue while the results obtained with \texttt{dynesty} and \texttt{ptemcee} are shown in orange and green, respectively.
All three posteriors produce consistent credible intervals. The means of pulses inevitably show degeneracy to some extent because of the overlap of SNRs between different components. However, the sampling problem is still relatively simplified compared with the case of modelling without ordering when all the different modes in posterior distribution have exactly the same shape.

\begin{figure}
    \centering
    \includegraphics[width=1.0\columnwidth]{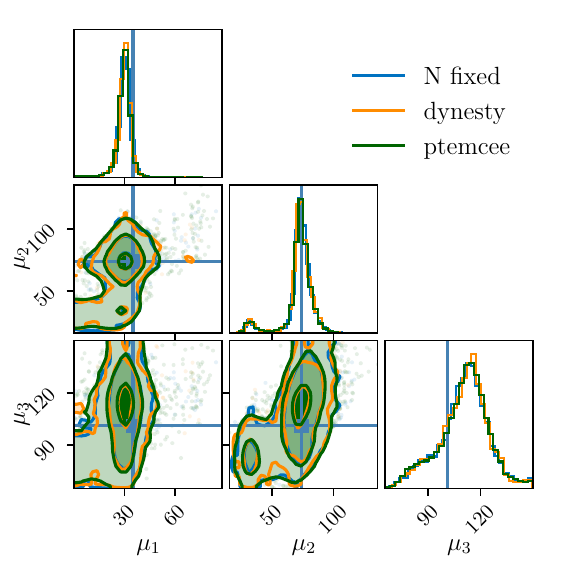}
    \caption{
    Corner plot of posterior samples of $N=3$ Gaussian pulses resorted by ascending means.
    The different shades show one, two, and three-sigma credible intervals.
    The ground truth, obtained with a fixed-$N$ \texttt{dynesty} run, is shown in blue.
    In orange we plot the results obtained using a transdimensional implementation of \texttt{dyensty} while green shows a transdimensional implementation of \texttt{ptemcee}.}
    \label{fig:processed_posterior_toy}
\end{figure}

For the sake of interpretability, we reorder the posterior samples by ascending means of Gaussian pulses and present the corner plot in Fig.~\ref{fig:processed_posterior_toy}. It is clear for the means of three pulses detected which are matched with the true values.

\subsection{GW150914}\label{sec: gw case}
We now apply transdimensional inference to reconstruct the signal from the first gravitational-wave observation GW150914. Strain data for GW150914 is accessed via the Gravitational Wave Open Science Center~\citep[GWOSC;][]{150914_strain_data}. We characterize the noise power spectral density using $\unit[128]{s}$ of data ahead of GW150914. We adopt sampling frequency as $\unit[2048]{Hz}$ with minimum and maximum cutoff frequencies of $\unit[20]{Hz}, \unit[896]{Hz}$. Following \texttt{BayesWave} \citep{Cornish_2015}, we assume that the source of gravitational waves is elliptically polarized so that the cross-polarized strain is completely determined by the plus-polarized strain:\footnote{For some bursting sources, it may be appropriate to adopt an unpolarized model so that $h_\times$ is modelled independently from $h_+$.}
\begin{equation}
    h_\times(f) = \epsilon h_+(f) e^{i\pi/2} .
\end{equation}
Here, $\epsilon \in [-1,1]$ is the ellipticity, which characterizes the polarization.
We fit the binary black hole signal GW150914 using a superposition of sine-Gaussian wavelets; see \cite{gw150914_bw}:
\begin{equation}
    \Psi(t | A, f_0, t_0, \, \phi) = Ae^{-(t-t_0)^2/\tau^2}\cos \big(2\pi f_0 (t-t_0)+\phi \big) 
\end{equation} 
with $\tau = Q/(2\pi f_0)$.
Here, $A$ is the amplitude, $\tau$ is the damping time, $Q$ is the quality factor, $t_0$ is the central time, $f_0$ is the central frequency, and $\phi$ is the phase offset. 
We approximate the SNR of a single wavelet using \citep{Cornish_2015}
\begin{equation}\label{eq:SNR}
    \text{SNR} \simeq \frac{A\sqrt{Q}}{(8\pi)^{1/4}f_0 S_n(f_0)}
\end{equation}
where $S_n$ is the noise power spectral density. 
The plus-polarized strain $h_+$ is the summation of several components
\begin{equation}
     h_+(t) = \sum_j \Psi(t | \text{SNR}_j, f_{j}, t_{j}, \tau_j, \phi_j).
\end{equation}

We adopt the following priors.
The distribution of non-ghost SNR $k\leq N$ are distributed according to Eqs.~\ref{eq:Beta}-\ref{eq:joint} for ordered statistics uniformly distributed on the interval $[0,30]$.\footnote{Since we are only interested in relative SNR, we ignore additional normalization constants in Eq.~\ref{eq:SNR}, treating the proportionality symbol in that equation as an equality.}
Ghost parameter SNRs $k>N$ are uniformly distributed on the interval $[0, 30]$.
The quality factor $Q$ is taken from a uniform distribution on the interval $[0.1 , 40]$, and $\phi$ follows a uniform distribution between 0 and $2\pi$ with periodic boundary conditions. 
We adopt a uniform prior for $f_{j}$ between $[\unit[20]{Hz}, \unit[512]{Hz}]$ and $t_{j}$ between $[\unit[-0.3]{s}, \unit[+0.2]{s}]$ around the trigger time.

\begin{figure}
    \centering
    \includegraphics[width=1\columnwidth]{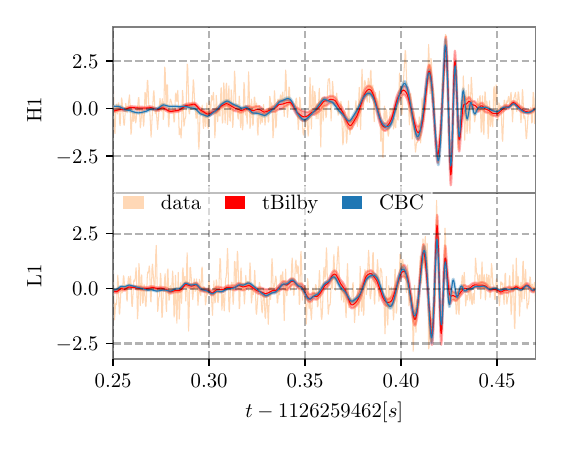}
    \caption{
    The reconstructed signal from GW150914.
    The red trace shows the whitened signal reconstructed using a transdimensional sine-Gaussian wavelet fit.
    The blue trace shows the maximum-likelihood template obtained with the \textsc{IMRPhenomXPHM} approximant.
    The whitened data is shown in peach.
    The top panel is for the H1 observatory in Hanford, WA while the bottom panel is for the L1 observatory in Livingston, LA.
    }
    \label{fig:150914}
\end{figure}

We analyze the LIGO--Virgo event GW150914 \citep{gw150914} using \texttt{dynesty} with the ghost parameter framework described above. 
We allow up to $N_\text{max}=8$ wavelets (45 total parameters).
We set up the analysis using rwalk method implemented in \bilby{} with nact=80 and nlive=2000. The sampling is converged in 4 days.
The reconstructed waveform is shown in Fig.~\ref{fig:150914} (red) alongside the whitened data (peach), and the compact binary coalescence (CBC) template fit shown as the blue curve.\footnote{The CBC fit is obtained using the waveform approximant \textsc{IMRPhenomXPHM} \citep{IMRPhenomXPHM}.} 
The top panel is for the LIGO Hanford Observatory (LHO) while the bottom panel is for the LIGO Livingston Observatory (LLO).
The wavelet fit produces a qualitatively similar reconstruction as the compact binary template fit.
Both fits recover the morphology of key features in the whitened data.

\begin{figure}
    \centering
    \includegraphics[width=1\columnwidth]{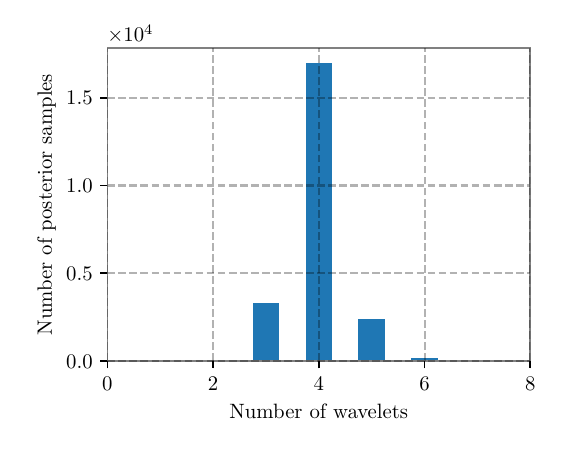}
    \caption{
    The model selection result for GW150914 wavelet fit. The ratio of number of samples at different $N$ represents the Bayes factor of different models.
    }
    \label{fig:150914_hist}
\end{figure}

In Fig. \ref{fig:150914_hist} we show the model selection result of GW150914 wavelet analysis. The posterior favours $N=4$ over $N=3$ and $N=5$ with Bayes factor of 5.13 and 7.13 respectively. Models of $N<3$ and $N>5$ wavelets are strongly disfavored. The wavelet fit produces a higher maximum likelihood than the template fit ($\Delta \ln {\cal L} = 6$) at the sample of $N=6$. Conditional posterior samples of the favored model $N=4$ have maximum $\ln {\cal L} = 306.5$ which is comparable to the maximum $\ln {\cal L} = 306.7$ of the template fit.
Since we expect the template derived from general relativity to fit the signal, we interpret this as evidence that the $N=6$ wavelet fit is beginning to overfit features in the noise.
In order to make further progress, it may be necessary to develop sampler settings that are better tuned for this transdimensional problem.
We plan to adjust the implementation of \texttt{dynesty} in \tbilby{} as a focus of future work.

\begin{figure*}
    \centering
    \includegraphics[width=0.9\textwidth]{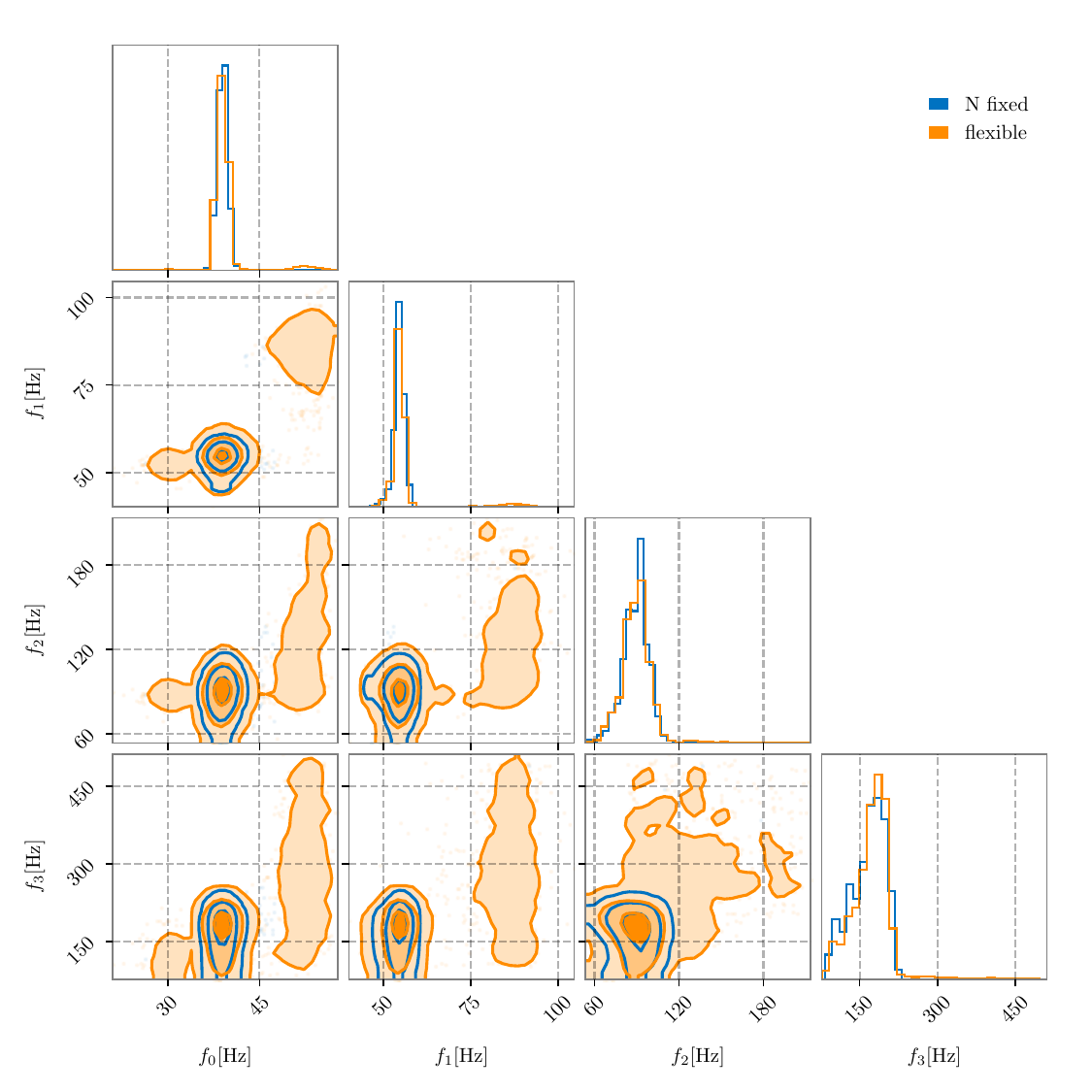}
    \caption{
    Posteriors for the frequencies of different sine-Gaussian wavelets fit to GW150914.
    The blue results are obtained with $N=4$ fixed.
    The orange shows posteriors of samples with $N=4$ from a transdimensional fit, which allows for any value of $N$.
    }
    \label{fig:150914_freq}
\end{figure*}

\begin{figure}
    \centering
    \includegraphics{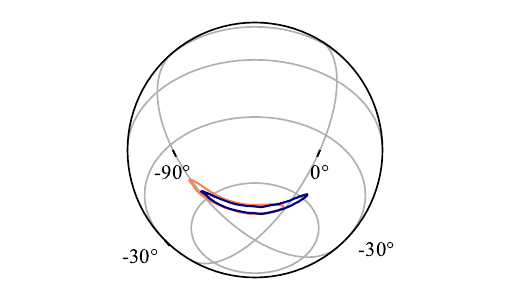}
    \caption{
    Sky map showing the 90 \% credible intervals for GW150914 recovered by a compact binary coalescence template fit (blue) and a transdimensional sine-Gaussian wavelet fit (orange).}
    \label{fig:skymap}
\end{figure}

In Fig. \ref{fig:150914_freq}, we show the posterior distributions for the frequencies of $N=4$ wavelets.
The blue posterior distributions are obtained with a fixed $N=4$ analysis using \bilby{}, while the orange results are obtained allowing for any value of $N$ using \tbilby. We reorder the samples from both analyses by ascending frequencies for the sake of interpretability.
In Fig.~\ref{fig:skymap} we show the sky localisation map for GW150914, where the blue curves are the 90\% credible intervals obtained using the \textsc{IMRPhenomXPHM} waveform approximant and the orange curves are obtained using our transdimensional sine-Gaussian wavelet fit.

As part of our validation, we also perform transdimensional inference on simulated data to validate our GW150914 reconstruction results. 
We inject 200 signals---drawn from our priors and consisting of different numbers of sine-Gaussians---to Gaussian noise, coloured to the LIGO O4 noise power spectral density.
See  Appendix \ref{appendix:injection} for complete details.

We find the sampler reconstructs 95\% of injections correctly with the true number of wavelets $N$. 
In the remaining 5\% of injections, the true value of $N$ is incorrectly ruled out, but for an interesting reason.
We determine that all of these injections include two wavelets, coincident in time, which the sampler fits using a single wavelet.
Thus, for example, an injection with $N=5$ is recovered with a strong preference for $N=4$.
In each case, the reconstructed waveform produces a good fit.
However, the sampler has difficulty, for example, jumping from an adequate $N=4$ fit to an also-adequate $N=5$ fit with two wavelets on top of each other.

To the extent that we care only about the adequacy of the reconstructed waveform, these wrong-$N$ recoveries are not especially concerning.
After all, the wavelet model is entirely phenomenological, and so there is no true value for $N$ in actual data.
On the other hand, it is desirable for model interpretability to eliminate this corner case.
A number of solutions may help including proximity priors (which prevent two wavelets from falling on top of each other in the first place) and custom jump proposals.
We leave this for future work.

\section{Discussion and conclusions}\label{Sec:discussion}
We introduce the \tbilby{} package that facilitates transdimensional inference calculations with \bilby.
Focusing, to start with, on time-domain models with a superposition of component functions, we provide examples where users can employ off-the-shelf samplers in Bilby to reconstruct signals with minimal alterations. 
The package includes example implementations of ghost parameters and order statistics, useful ingredients for this class of transdimensional problems.
We show how \tbilby{} can be used to perform a minimum-assumption fit of GW150914 with sine-Gaussian wavelets as in \cite{gw150914_bw}.

For future work, we propose to improve the efficiency of \tbilby{} through the use of more finely tuned samplers, designed for specific classes of problems of interest in gravitational-wave astronomy.
Thanks to the modular design of \bilby, it is relatively easy to experiment with different options.
While we find that \texttt{dynesty} produces well-converged fits to GW150914 for $N_\text{max}\leq8$, we do not obtain reliable fits with \texttt{ptemcee} in a decent sampling time---at least using the default settings.
Our work highlights the potential for carrying out transdimensional inference with nested sampling; see, e.g., \cite{Brewer}.

We see this paper as the first step in a broader program to facilitate transdimensional inference with \bilby---in gravitational-wave astronomy and other contexts.
We highlight a few priorities.
First, as evidenced by work done with \texttt{BayesLine} \citep{BayesLine}, transdimensional inference is a powerful tool for modelling the noise in gravitational-wave observatories; see also \cite{Gupta}.
Noise modelling naturally lends itself to transdimensional models because the noise power spectral density can be characterised by some fiducial shape plus a variable number of spectral features superposed on top.
Transdimensional models can be used to obtain smooth fits of the noise power spectral density while characterizing instrumental lines and other features, enabling us to study the evolution of these features over the course of an observing run. We are also excited about the application of transdimensional sampling to model potential glitches simultaneously with compact binary signals \citep{Chatziioannou_cbc_glitch, Hourihane_2022}. This work may help astronomers to better interpret gravitational-wave events with potential data quality problems \citep{PhysRevD.106.104017}.
A comprehensive study of noise modelling using \tbilby{} will be detailed in a forthcoming publication.

Second, we envision extending \tbilby{} to build more flexible models describing the population properties of binary black holes and neutron stars; see, e.g., \cite{Toubiana}.
For example, one may wish to model the distribution of primary black hole mass mass distribution with a variable number of peaks and troughs. 
Recent studies have highlighted the usefulness of flexible models to identify structure that might be missing from astrophysically inspired phenomenological models; see, e.g., \cite{Tiwari21,Edelman22,Edelman23}.

Finally, we propose to develop \tbilby{} for applications beyond terrestrial gravitational-wave observatories.
For example, pulsar timing measurements, which can be used to measure nanohertz gravitational-waves \citep{NANOGrav,EPTA,PPTA,CPTA}, rely on measurements of the time of arrival of arbitrarily shaped radio pulses.
By modelling these pulses using a superposition of component functions, it is sometimes possible to identify and account for aberrant behaviour in the pulsar evolution, ultimately improving sensitivity for gravitational-wave searches \citep{Nathan_2023}. 
Transdimensional models may prove useful determining the number of component functions used in these fits.
Of course, this is just one example.
It is our hope that the \tbilby{} package will facilitate the development of numerous transdimensional models for physics and astronomy.

\begin{acknowledgments}
This work is supported through Australian Research Council (ARC) Centres of Excellence CE170100004, CE230100016, Discovery Projects DP220101610 and DP230103088, and LIEF Project LE210100002.
T. A. C. receives support from the Australian Government Research Training Program. 
The authors are grateful for for computational resources provided by the LIGO Laboratory computing cluster at California Institute of Technology supported by National Science Foundation Grants PHY-0757058 and PHY-0823459, and the Ngarrgu Tindebeek / OzSTAR Australian national facility at Swinburne University of Technology.
LIGO was constructed by the California Institute of Technology and Massachusetts Institute of Technology with funding from the National Science Foundation and operates under cooperative agreement PHY-1764464. This paper carries LIGO Document Number LIGO-P2400105.
\end{acknowledgments}

\appendix

\section{Code design}\label{appendix:tbilby_package}
The objective of \tbilby{} is to provide a comprehensive toolkit for handling transdimensional sampling. 
The \tbilby{} package offers flexibility and automation. 
As outlined in this paper, the development of \tbilby{} is part of a long-term project with multiple goals. At present, we have constrained the package to a set of essential tools and examples.
\tbilby{}'s design philosophy closely aligns with that of \bilby, emphasizing open-source code, modularity, generality, and usability \citep{bilby}. Based on the ideas and infrastructure of \bilby, \tbilby{} ensures a relatively smooth user experience, particularly for experienced users. 
Furthermore, we reinforce this philosophy by mandating that the sole requirement for \tbilby{} is an installation of \bilby.

The structure of \tbilby{} closely mirrors that of \bilby, with the core module including \texttt{base}, \texttt{prior}, and \texttt{sampler} modules, alongside an additional folder dedicated to examples. 
The \texttt{base} module contains fundamental functionality for constructing transdimensional models and defining transdimensional priors. 
The \texttt{prior} folder houses priors intended for transdimensional sampling, while the \texttt{sampler} module facilitates support for transdimensional samplers.

The key building block of a transdimensional model in \tbilby{} is the \textit{transdimensional parameter}, which refers to a parameter of a component function that has multiple ``orders'' (in this language, each sine-Gaussian is a different order). 
Another fundamental concept is the \textit{transdimensional prior}, which constitutes a set of priors related to a transdimensional parameter and which is attached to the parameter's order. 
Transdimensional models with proximity priors employ conditional statements.
These two elements serve as the basic building blocks. 

For practical purposes, transdimensional priors in \tbilby{} are categorized into four types: (i) transdimensional nested conditional priors, (ii) transdimensional conditional priors, (iii) conditional priors, and (iv) unconditional priors.\footnote{Examples employing each of the prior types can be found at the \texttt{git} repository.}
Transdimensional nested conditional priors are defined by their dependence on previously sampled parameters of the same component function. 
If we assume that the current order being sampled is $n$, these priors depend on parameters of orders $n-1, n-2,$ etc.

Transdimensional conditional priors, on the other hand, are dependent on parameters from all sampled orders of a component function, denoted by $k, k-1,$ etc.
Conditional priors rely on a set of non-transdimensional parameters, whereas unconditional priors are independent of other parameters.
The most general prior may combine elements of all these types, except for the last type, which by definition is an independent prior. 
In this framework, the most general form of a prior for transdimensional parameter $\rho_n$ is:
\[ \pi\left(\rho_{n} | \rho_{n-1},  \ldots \phi_{n-1}, \ldots \zeta_{k}, \zeta_{k-1}, \ldots \Lambda \right). \]
The variable $\phi$ represents another set of transdimensional parameters of the same order as the component function so that $\rho_n$ does not depend on $\phi_{m \geq n}$. 
Meanwhile, $\zeta$ signifies another set of transdimensional parameters that depend on all available orders of the component function. 
Finally, $\Lambda$ are parameters that may or may not be part of the component function parameters.

By allowing for the definition of conditional transdimensional priors, users can uniquely specify priors for each transdimensional parameter.
Practically, this involves defining a class that inherits from a predefined transdimensional prior class and implementing an abstract function to define the mathematical relation between the conditional parameters and prior properties (this is a generalization of \bilby's  condition function, which is required when defining a conditional prior). 

Facilitating such versatility and control over the priors allows users to gain flexibility in manipulating the prior distribution to suit their specific needs.
The flexibility of \tbilby{}'s extends further, enabling the construction of function superposition, each potentially comprising a different number of component functions. 
For instance, the LIGO noise power spectral density may be represented as a combination of several power law functions along with multiple Cauchy-like functions, addressing distinct spectral characteristics \citep{BayesLine}.
Furthermore, \tbilby{} offers supplementary tools for removing ghost parameters and generating relevant corner plots, thereby simplifying the analysis of component functions and individual transdimensional parameters.

\section{Ghost parameters}\label{appendix:ghost parameter}
The method outlined here is similar to \cite{Liu_2023} who performed transdimensional inference using \bilby{} for gravitational-wave lensing study. In the ghost parameter framework, we introduce extra parameters that do not actually change the likelihood, and therefore do not change the posteriors for the original parameters---as long as the ghost-parameter prior is correctly normalized. 
For example, we consider the situation in Section~\ref{sec:pulses} when $N = 3$.
The signal is only determined by $\theta_{k\leq3}$ while $\theta'_{k>3}$ represents the ghost parameters. 
In this case, the posterior is
\begin{equation}
    p(\theta_{k\leq3},\theta_{k>3}, N = 3|d) = p(\theta_{k\leq3},\theta_{k>3}|d, N = 3)p(N = 3|d).
\end{equation}
The conditional posterior given $N$ can be written as
\begin{equation}\label{eq:ghost condition posterior}
    p(\theta_{k\leq3},\theta_{k>3}|d, N = 3) = \frac{\mathcal{L}(d|\theta_{k\leq3})\pi(\theta_{k\leq3})\pi(\theta_{k>3}|\theta_{k\leq3})}{p(N=3|d)/\pi(N=3)}.
\end{equation}
As the priors for the extra parameters are properly normalized by definition, i.e., 
\begin{equation}
\int p(\theta'_{k>3}|\theta_{k\leq3})d\theta'_{k>3} = 1, 
\end{equation}
the marginalized posterior for $\theta_{k\leq3}$ is equivalent to the case where there are no extra parameters:
\begin{equation}
\begin{aligned}
    p(\theta_{k\leq3}|d, N = 3) \propto & \int \mathcal{L}(d|\theta_{k\leq3})\pi(\theta_{k\leq3})\pi(\theta_{k>3}|\theta_{k\leq3})d\theta_{k>3} \\
    \propto& \mathcal{L}(d|\theta_{k\leq3})\pi(\theta_{k\leq3}).
\end{aligned}
\end{equation}
Now we take a look at the denominator of Eq. \ref{eq:ghost condition posterior}. It is actually the marginal likelihood of $N$ in transdimensional sampling:
\begin{equation}
    \mathcal{L}(d|N=3)=p(N=3|d)/\pi(N=3).
\end{equation}
Meanwhile, we note it is essentially a normalization factor, so the expression can be also written as
\begin{equation}
\begin{aligned}
    \mathcal{L}(d|N=3) = & \int \mathcal{L}(d|N=3, \theta_{k\leq3}) \times \\
    &\pi(\theta_{k\leq3}) \pi(\theta'_{k>3}|\theta_{k\leq3}) d\theta_{k\leq3} d\theta_{k>3}\\
    =& \int \mathcal{L}(d|\theta_{k\leq3})\pi(\theta_{k\leq3}) d\theta_{k\leq3}\\
    =& \mathcal{Z}_{N=3}.
\end{aligned}
\end{equation}
We make use of the fact that the priors for ghost parameters are properly normalized again.

So the model selection result of our transdimesional problem with ghost parameters is valid regardless of the inclusion of ghost parameters as the likelihood $\mathcal{L}(d|N=3)$ is correctly defined as the case without the implementation of ghost parameters.

As a comparison, the detailed balance equations of traditional reversible jump Markov chain Monte Carlo without ghost parameters is written as
\begin{equation}\label{eq:w/o ghost}
    p(\theta_{k\leq3}|d)q(\theta'_{k\leq3}, \theta'_{k>3}) = p(\theta'_{k\leq3}, \theta'_{k>3}|d) q(\theta_{k\leq3})\alpha,
\end{equation}
where $p(\theta|d)$ is the target distribution, i.e., posteriors in Bayesian inference, $q(\theta)$ is the proposal for samples in Markov chain Monte Carlo sampling and $\alpha$ is the acceptance probability. 
This makes use of the trade-off between higher dimension proposals in the left-hand side and higher dimension posteriors in the right-hand side.

With the implementation of ghost parameters, we artificially add extra dimensions for posteriors and proposals in both side with the detailed balance equation written as
\begin{equation}
    p(\theta_{k\leq3},\theta_{k>3}|d)q(\theta'_{k\leq3}, \theta'_{k>3}) = p(\theta'_{k\leq3}, \theta'_{k>3}|d) q(\theta_{k\leq3}, \theta_{k>3})\alpha.
\end{equation}

As we show above, in the case where $\theta_{k>3}$ are not used in the evaluation of the likelihood, the posteriors could be written as two independent parts
\begin{equation}
    p(\theta_{k\leq3},\theta_{k>3}|d) = p(\theta_{k\leq3}|d)\times \pi(\theta'_{k>3}|\theta_{k\leq3}).
\end{equation}
Thus, if we choose a proper reversible proposal distribution to make
\begin{equation}
\begin{aligned}
    q(\theta'_{k\leq3}, \theta'_{k>3}) = &q(\theta'_{k\leq3})q(\theta'_{k>3}|\theta'_{k\leq3}) \nonumber\\
    = & q(\theta'_{k\leq3})\pi(\theta_{k>3}|\theta_{k\leq3}),
\end{aligned}
\end{equation}
then the detailed balance can be written as
\begin{equation}
\begin{aligned}
    &p(\theta_{k\leq3}|d) \pi(\theta_{k>3}|\theta_{k\leq3})q(\theta'_{k\leq3}, \theta'_{k>3}) = \\&p(\theta'_{k\leq3}, \theta'_{k>3}|d) q(\theta'_{k\leq3})\pi(\theta_{k>3}|\theta_{k\leq3})\alpha .
\end{aligned}
\end{equation}
This reduces to Eq.\ref{eq:w/o ghost} where we do not implement ghost parameters.
In fact, it is not necessary to set up special proposals for ghost parameters. 
With arbitrary proposal distributions, the sampling result with the implementation of ghost parameters will always be consistent with the situation without ghost parameters as the statistical average of acceptance rate $\alpha$ over the entire parameters space.

\section{The order statistics of a uniform distribution}\label{appendix:order stats}
In this Appendix, we describe the order statistics of a uniform distribution.
This formalism imposes an order of different components while maintaining the uniform distribution of the unordered statitics.
From \cite{order_stats}, if one draws $N$ samples from a uniform distribution on the interval $[0,1]$ and arranges them in ascending order, for $i<j$, the joint probability density function of the $i^\text{th}$ and $j^\text{th}$ samples $(u,v)$ can be shown to be
\begin{equation}\label{eq:order_2d}
    f(u,v|N)=N!\frac{u^{i-1}}{(i-1)!}\frac{(v-u)^{j-i-1}}{(j-i-1)!}\frac{(1-v)^{n-j}}{(n-j)!}.
\end{equation}

Since the $u$ variables are ordered by construction, the conditional prior for $u_i$ depends on $u_{i-1}$. But it does not depend on $u_j$ explicitly for $j<i-1$. 
It is already taken care of that $u_i<u_{i-2}$ because $u_{i-1}<u_{i-2}$. 
We can use the fact
\begin{equation}
    f(u_i|u_{i-1},u_{i-2},...,u_0,N)=f(u_i|u_{i-1},N),
\end{equation}
to sample $u_i$ in order given Eq.~\ref{eq:order_2d}.
The joint density of all N variables is constant
\begin{equation}\label{eq:joint_order_stats}
    f(u_1,u_2,...,u_N|N)=N! .
\end{equation}
The normalization factor accounts for the decreased parameter space by the factor of number of permutations $N!$. 

The marginalized likelihoods, i.e., evidence values of those two kinds of priors, are the same, which indicates no additional biases are introduced to our transdimensional sampling by the order statistics uniform prior.
The overall evidence using uniform priors $F(u_1,u_2,...,u_N)$ without ordering can be calculated by
\begin{equation}
    \mathcal{Z} =\int \mathcal{L}(d|u_1,u_2,...,u_N)F(u_1,u_2,...,u_N)du.
\end{equation}
This integral can be refactored to be the summation of integrals in the parameter spaces by different permutations of $u_i$; i.e.,
\begin{equation}
\begin{aligned}
    \mathcal{Z} =&\int \mathcal{L}(d|u_1,u_2,...,u_N)F(u_1,u_2,...,u_N)du\\
    =&\int_{u_1>u_2>u_3>...>u_N} \mathcal{L}(d|u_1,u_2,...,u_N)F(u_1,u_2,...,u_N)du + \\
    &\int_{u_2>u_1>u_3>...>u_N} \mathcal{L}(d|u_1,u_2,...,u_N)F(u_1,u_2,...,u_N)du + ...
\end{aligned}
\end{equation}
Each integral of parameter space by one permutation will lead to the same value, i.e., $\mathcal{Z}/N!$.

For the evidence obtained with the priors, $f(u_1,u_2,...,u_N|N)$ is distributed according to the order statistics of a uniform distribution, the integral happens in one parameter subspace, which follows the permutation. 
However, the density of the priors increase by a factor of $N!$, as we show in Eq.\ref{eq:joint_order_stats}, so that the evidence $\mathcal{Z}'$ remains the same
\begin{equation}\label{eq:order_evidence}
\begin{aligned}
    \mathcal{Z}' &=\int_{u_1>u_2>u_3>...>u_N} \mathcal{L}(d|u_1,u_2,...,u_N)f(u_1,u_2,...,u_N|N)du \\
    &=N!\times\int_{u_1>u_2>u_3>...>u_N} \mathcal{L}(d|u_1,u_2,...,u_N)F(u_1,u_2,...,u_N)du\\
    &=N!\frac{\mathcal{Z}}{N!}\\
    &=\mathcal{Z}
\end{aligned}
\end{equation}

Finally, we note that one might be tempted to design a prior where $u_1$ is uniform on the interval $[0,1]$, $u_2$ is uniform on the interval $[0, u_1]$, and so on.
Naively, one might expect this to reproduce the distribution described above, but it does not.
The prior for $\mu_k$ rails against 1 for large values of $k$ and the distribution is not equivalent to what one obtains by drawing $N$ uniform numbers and then ordering them after the fact.

\section{Alternative ordering schemes}\label{alternative}
In this Appendix, we explore alternative ordering schemes.
While our examples above employ ordering in SNR, one can, in principle, order in any component function parameter.
Mathematically, the prior is the same no matter how one chooses to order the parameters, and so the ordering scheme is important only to ensure a converged result and for interpretability.

For example, let us reanalyse the superposition of Gaussian pulses but ordering by center time $\mu$. 
Let us denote SNR-ordered parameters with subscript numbers $(\mu_1,\mu_2,..., \rho_1,...)$.
We contrast this with $\mu$-ordered parameters with subscript letters: $(\mu_A,\mu_B,..., \rho_A,...)$.
The likelihood of the data given different ordering schemes is the same since different ordering schemes just relabel the parameters:
\begin{equation}
    \mathcal{L}(d|\mu_1,\mu_2,..., \rho_1,...) = \mathcal{L}(d|\mu_A,\mu_B,..., \rho_A,...).
\end{equation}

\begin{figure*}
    \centering
    \includegraphics[width=0.9\textwidth]{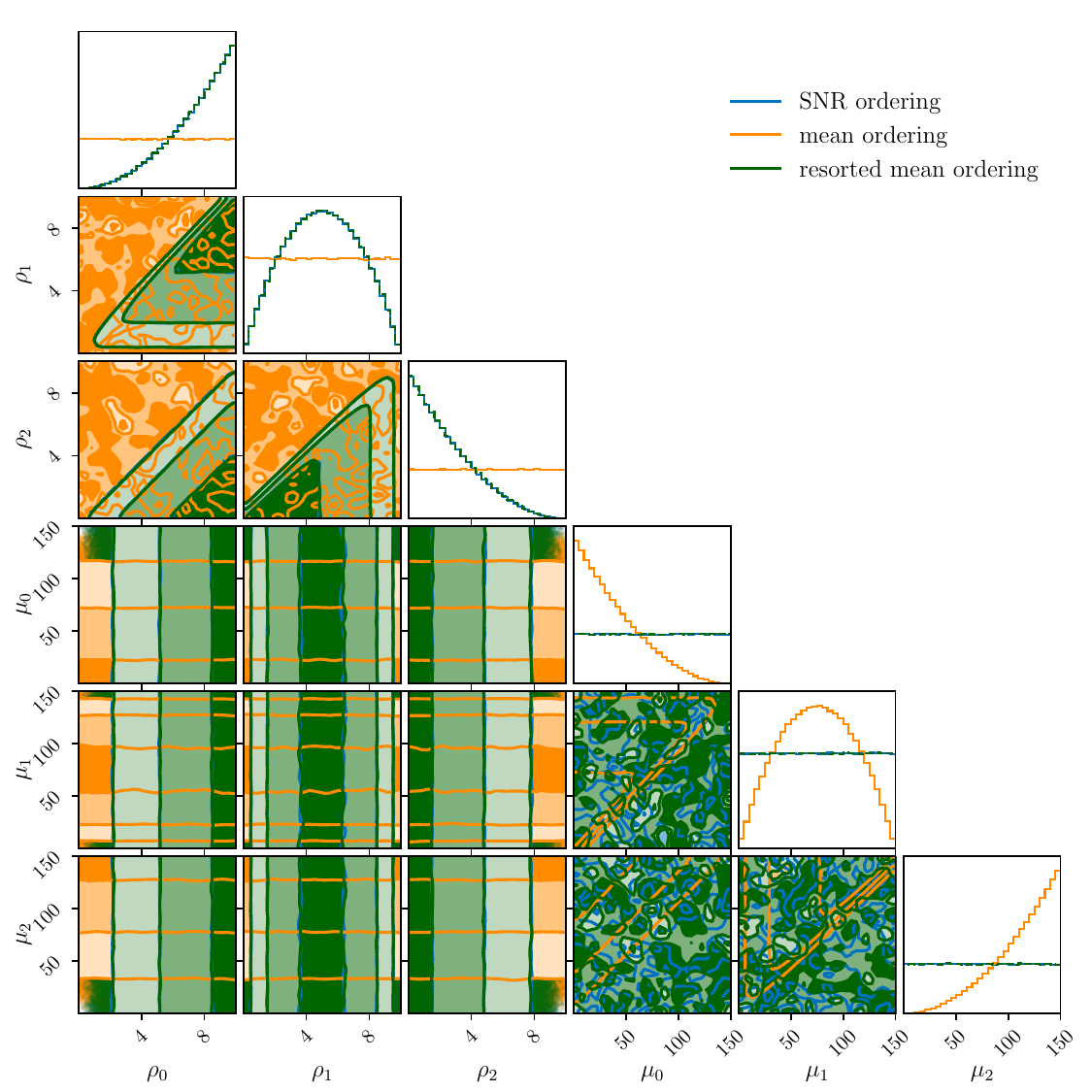}
    \caption{
    A corner plot shows the distribution of different order schemes. Blue is ordering by SNRs and uniform means priors while orange is ordering by means and uniform SNRs. We reconstruct the distribution in green by resorting the samples from the distribution in orange by descending SNRs. Note the blue almost completely overlaps the green as we expect.
    }
    \label{fig:priors_ordering_schemes_conversion}
\end{figure*}

Likewise, ordering schemes do not change the shape of the prior distribution; though, they increase the overall probability density by $N!$ compared to the unordered prior.
This compensates the decrease in parameter space that occurs when completely degenerate modes are eliminated with ordering. 
Thus, different ordering schemes lead to the same prior values.
For example, if we draw samples from the SNR-ordered prior
\begin{equation}
   \pi(\mu_1,\mu_2,..., \rho_1,...),
\end{equation}
and convert the samples to $\mu$-ordering, they will match samples drawn from the $\mu$-ordered prior:
\begin{align}
    \pi(\mu_A,\mu_A,..., \rho_A,...).
\end{align}
We demonstrate this in Fig.~\ref{fig:priors_ordering_schemes_conversion}. 

\begin{figure*}
    \centering
    \includegraphics[width=0.9\textwidth]{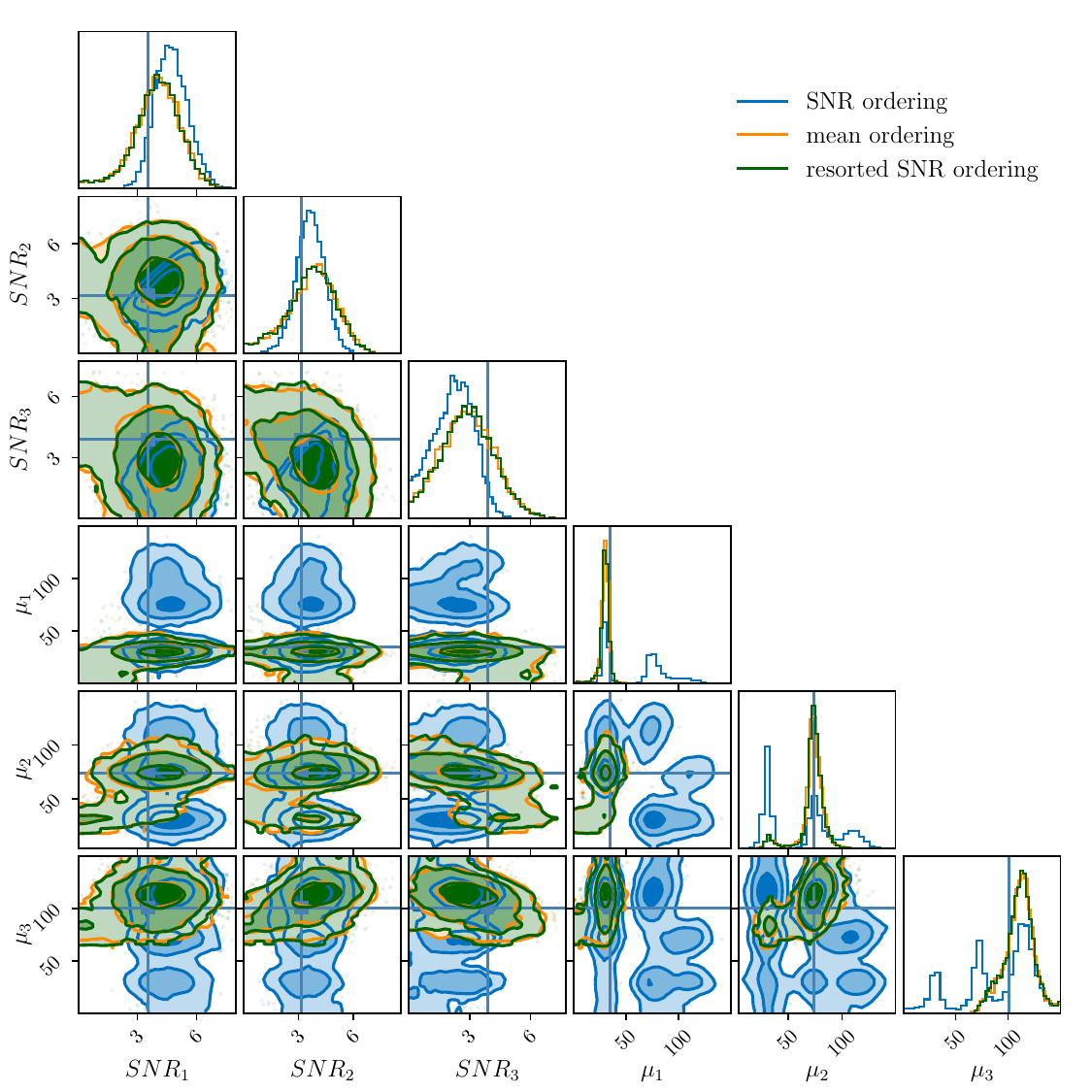}
    \caption{
    Posteriors of Gaussian pulses using different order schemes when N fixed at 3. Blue is using SNR ordering priors while orange is using mean ordering priors. We reconstruct the distribution in green by resorting the samples from distribution in blue by ascending means.
    }
    \label{fig:posteriors_ordering_schemes_conversion}
\end{figure*}

Since the choice of ordering does not affect the likelihood or the prior, the posteriors for different ordering schemes are mathematically equivalent. 
In Fig.~\ref{fig:posteriors_ordering_schemes_conversion}, we relabel the posterior samples from SNR ordering priors by ascending $\mu$.
The resultant distribution (green) is consistent with the posterior by $\mu$-ordered priors (orange). 
The Bayes factors are also consistent; see Table~\ref{tab:BF_ordering_schemes}. 

As noted above, the fact that different ordering schemes are mathematically equivalent does not imply they are all equally useful in practice.
To illustrate this, we revisit the reconstruction of GW150914, but this time employing a  frequency-ordered prior. 
The sampling process gets stuck at N=2 with a natural log maximum likelihood value $\sim100$ less than we obtain with SNR-ordered runs, which favor samples with $N=4$.
The sampler successfully captures the first two most significant features, i.e., the components with highest SNRs, which have frequencies around $\unit[85]{Hz}$ and $\unit[140]{Hz}$. 
This result is consistent with what we find when we analyse GW150914 data with $N=2$.
However, the central frequency of the third component is $\lesssim\unit[140]{Hz}$, which is unlikely to be identified in the frequency-ordered scheme because the second wavelet to be identified already has $f>\unit[140]{Hz}$. 
Therefore it is hard for the sampler to escape from the $N=2$ local likelihood maximum.

\begin{table}
\centering
 \begin{tabular}{|c| c| c|} 
 \hline
 &\multicolumn{2}{c|}{\textbf{ln Bayes factor}}\\
 \hline
 N & SNR ordering & mean ordering\\
 \hline
 1 &  8.12&  8.16 \\
 \hline
 2 & 13.03 & 13.08 \\
 \hline
 3 & 14.33 & 14.01 \\
 \hline
 4 & 13.94 & 13.91 \\
 \hline
 5 & 13.06 & 13.13 \\
 \hline
 6 & 11.82 & 12.13 \\
 \hline
 \end{tabular}
 \caption{
 Natural log Bayes factors relative to noise obtained with assumptions of different number of Gaussian pulses and different ordering schemes in priors.
 }
 \label{tab:BF_ordering_schemes}
\end{table}

\section{Sine-Gaussain injection study}\label{appendix:injection}
We perform an injection study to validate our transdimensional result of wavelets fit. The distribution of our simulated parameters is equivalent to the prior used for our GW150914 analysis with some exceptions. The distribution of non-ghost SNR $k\leq N$ are distributed according to Eqs.~\ref{eq:Beta}-\ref{eq:joint} for ordered statistics uniformly distributed on the interval $[10,20]$ where the bulk of GW150914 posterior support lies.
Ghost parameter SNRs $k>N$ are uniformly distributed on the interval $[10, 20]$.
The quality factor $Q$ is taken from a uniform distribution on the interval $[2 , 40]$. 
We exclude the small $Q<2$ region deliberately because it is hard to fit the delta function like signal for our sampler currently. 
The prior for the number of wavelets is uniform on $[3, 6]$.
We use the standard LIGO O4 noise power spectral density. Also, to limit computational costs, we only consider a single detector. 
We expect that multiple detectors would produce similar results. 

\begin{figure*}
    \centering
    \includegraphics[width=0.9\textwidth]{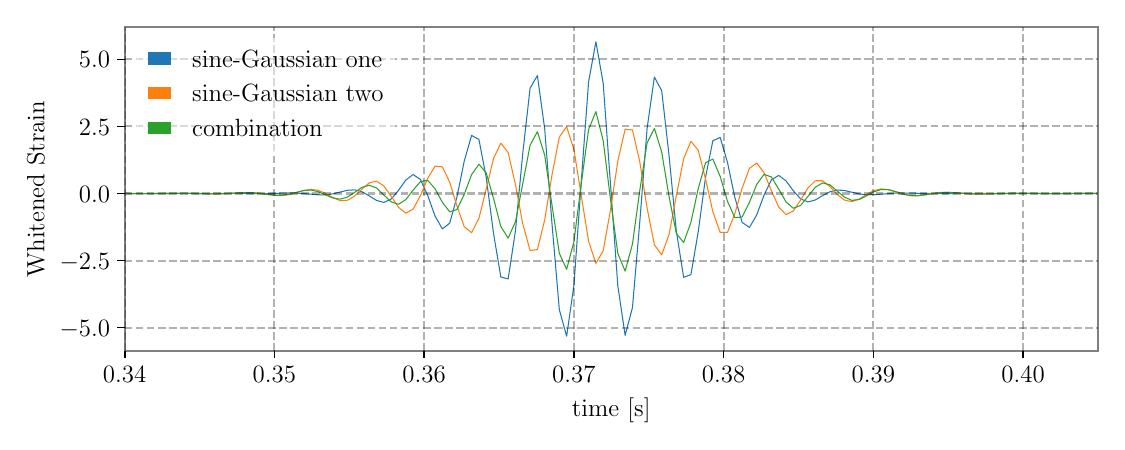}
    \caption{
    Example of the failure when the sampler can't identify the number of wavelets correctly. Blue and orange represents the two separate sine-Gaussians in the simulated data. Green is the combination of these two.
    }
    \label{fig:overlap_example}
\end{figure*}

\begin{figure*}
    \centering
    \includegraphics[width=0.9\textwidth]{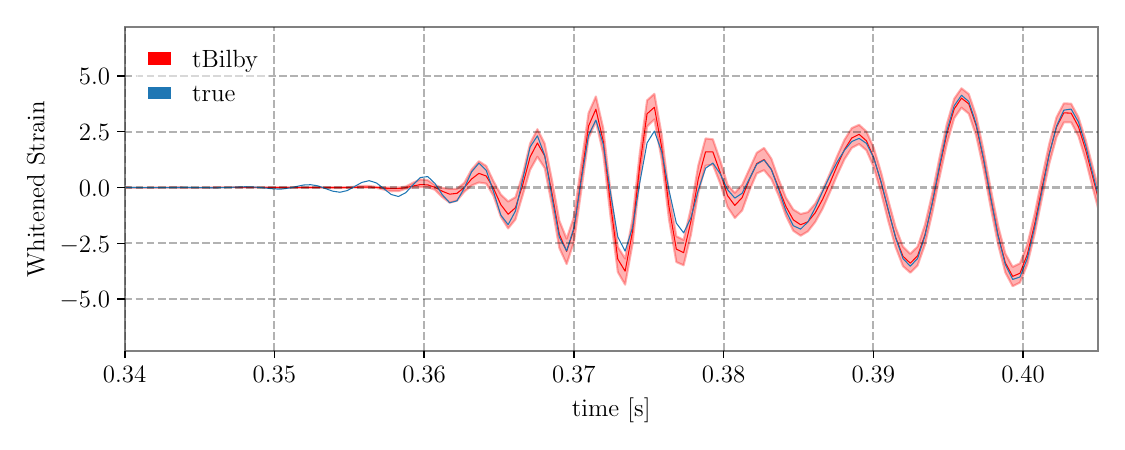}
    \caption{
    Example of the reconstructed waveform when the sampler can't identify the number of wavelets correctly. Blue is the injected signal and red is the fitting result.
    }
    \label{fig:overlap_fit_result}
\end{figure*}

We find that 190 out of 200 injections the true value of $N$ is strongly preferred ($\gtrsim 99\%$ credibility). 
In the cases where the sampler fails to correctly identify the injected number of wavelets, there is a strong preference for $N_\text{true}-1$. 
Upon further investigation, we determine that all 10 of these injections fit the same pattern. 
There are two wavelets, overlapping in time, which can be approximately described by one wavelet (see Fig. \ref{fig:overlap_example} for example).
However, the reconstructed waveform produces a reasonable match to the injected signal; See Fig.~\ref{fig:overlap_fit_result}.

Various solutions may be possible to eliminate this corner case.
One option is to employ proximity priors, which prevent two wavelets from falling on top of each other so as to produce a signal that looks like one wavelet.
Since wavelet models are  phenomenological, it is worth designing them to be minimise the challenges for the sampler.
Another possibility is to employ custom jumps so that the sampler more efficiently explores the possibility of replacing one wavelet with a superposition of two.

\bibliography{refs}{}
\bibliographystyle{aasjournal}

\end{document}